# Interface defect-assisted phonon scattering of hot carriers in graphene

Sergey G. Menabde,[1] Hyunwoo Cho,[1] and Namkyoo Park[1,*]

[1]*Dept. of Electrical and Computer Engineering, Seoul National University, Seoul, 08826, Korea*
[*]nkpark@snu.ac.kr

The broadband and ultrafast photoresponse of graphene has been extensively studied in recent years, although the photoexcited carrier dynamics is still far from being completely understood. Different experimental approaches imply either one of two fundamentally different scattering mechanisms for hot electrons. One is high-energy optical phonons, while the other is disorder-driven supercollisions with acoustic phonons. However, the concurrent relaxation via both optical and acoustic phonons has not been considered so far, hindering the interpretation of different experiments within a unified framework. Here we expand the optical phonon-mediated cooling model, to include electron scattering with the acoustic phonons. By assuming the enhancement of electron-acoustic phonon supercollisions from the localized defect at the photothermoelectric current-generating interface, we provide a broader perspective to the ultrafast photoresponse of graphene, highlighting the previously overlooked effect of the interface for cooling dynamics. We show that the transient photothermoelectric response, which has been attributed exclusively to supercollisions, can be successfully explained without rejecting the established optical phonon relaxation pathway, demonstrating that the two cooling mechanisms are not mutually exclusive but complement each other.

## I. INTRODUCTION

The scope of light-matter interaction has been significantly enriched with the discovery of graphene in the middle 2000s. An unusual gapless and linear band structure of electrons behaving as massless Dirac fermions is accountable for an ultrabroadband and nontrivial electrodynamic response from the single layer of carbon atoms [1]. Nonetheless, the exact cooling dynamics of the photoexcited carriers in graphene is not yet completely understood. So far, two contradicting models for the main cooling mechanism have been suggested: the electron-optical phonon [2-5] scattering (OP model), and the disorder-driven electron-acoustic phonon scattering, or supercollisions, which alleviates the strict momentum conservation [6] (SC model). These models suggest fundamentally different pathways of carrier relaxation in graphene, however each has its own experimental confirmation.

The first proposed model is based on electron-optical phonon scattering only [5,7,8], and referred to as the OP model. This analytical model is supported by ample experimental data obtained with time-resolved Raman spectroscopy [9] or time- and angle-resolved photoemission study [10] of graphite, ultrafast photoluminescence [11] or optical spectroscopy of graphene with optical pump-probe [5,12] and optical-pump terahertz-probe [13,14], as well as similar study on graphite [15]. In this model it is generally adopted that, the optical phonon-mediated cooling is governed by the coupling strength between the carriers and the phonons [4,5], and therefore the relaxation time is a function of the lattice temperature, the pump fluence, and the Fermi level. The OP model, or its refined versions such as microscopic approach [14], successfully predicts a wide range of experimentally observed carrier relaxation time from a few to several hundred picoseconds [8,13,14] with phonon lifetime of 1~2 ps [5,9,14,16-18]. Together with the most recent study by Yang *et al*. [19], which details the transient process, these past experimental results strongly support that the optical phonon bath is an efficient energy drain path for hot carriers. To compare, the recently suggested SC model emphasizes the exclusive role of disorder-assisted scattering [6]. The SC cooling model has been proved to provide a good fit with the transient photothermoelectric (PTE) photoresponse in graphene, for example, at the interface between mono- and bi-layer graphene [20], at the *p-n* junction formed by non-uniform doping [21,22], or at the graphene-metal contact [23,24].

However, we note that overall, neither of the OP and SC models provide a complete theoretical platform for every case. For example, the OP model fails to predict the PTE photoresponse in graphene at low temperatures [21,25], and the SC model's prediction for disorder-driven-only cooling contradicts with the disorder-independent, very similar relaxation time obtained by spectroscopic measurements made on graphene with a different number of impurities and Fermi level [5,14].

In this study, we address the discrepancy between the existing models of hot carrier cooling in graphene. Specifically we note that, the spectroscopic studies on uniform graphene strongly suggest the major role of optical phonons, but their contribution has been completely ignored by the studies focused on the disorder-attributed PTE photoresponse in graphene *p-n* junction (or metal contact). We show that this inconsistency could be ameliorated by including the role of the current generating interface in the OP model as a localized defect, which enhances the supercollision-mediated carrier relaxation (Fig. 1). Our model of interface defect-assisted phonon scattering thus offers a unified theoretical framework for the cooling dynamics of photoexcited carriers in graphene, for both the PTE photocurrent and the spectroscopic experiments. Verification of the model is provided by comparing the result with the published experimental data: in terms of the transient PTE photocurrent decay (including the nonlinear power dependence), the peculiar PTE response at different doping, the relaxation rates, and the analytical heat transfer rates $\Gamma$ given by acoustic and optical phonon scattering, all with excellent agreement.

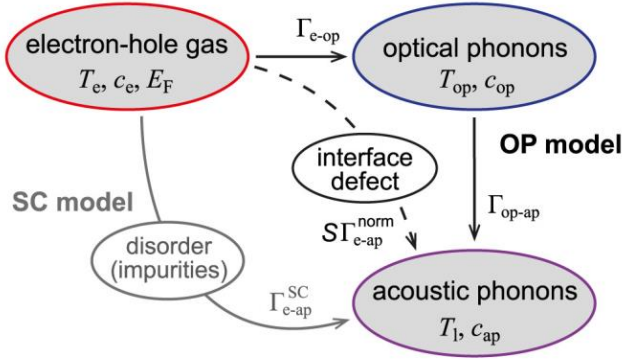

FIG. 1. A schematic of the three-temperature formalism. Net cooling power flows are shown as arrows. The SC model assumes only the disorder-assisted scattering of hot carriers by acoustic phonons, while the OP scenario assumes momentum-conserved relaxation via the optical (and acoustic) phonons. The interface-assisted supercollision enhancement factor $S$ is set to be zero in uniform graphene, since very weak normal scattering with the acoustic phonons (dashed arrow) is usually ignored in the OP model.

## II. CARRIER COOLING DYNAMICS IN PRESENCE OF INTERFACE DEFECT

We start with a three-temperature approach [10], selectively considering different sets of parameters for the OP and SC scenarios (Fig. 1). In this formalism, graphene is treated as a threefold system consisting of thermalized subsystems with corresponding temperatures $T$: the electron-hole gas ($T_e$) being in equilibrium and following the Fermi-Dirac distribution [26-29], optical phonons ($T_{op}$), and the acoustic phonons or lattice ($T_l$). Temporal evolution of the system can then be described by the coupled rate equations:

$$\begin{cases} \dfrac{dT_e}{dt} = \dfrac{H - \Gamma_{e\text{-}op} - \Gamma_{e\text{-}ap}}{c_e} \\ \dfrac{dT_{op}}{dt} = \dfrac{\Gamma_{e\text{-}op} - \Gamma_{op\text{-}ap}}{c_{op}} \\ \dfrac{dT_l}{dt} = \dfrac{\Gamma_{e\text{-}ap} + \Gamma_{op\text{-}ap}}{c_{ap}} \end{cases} \quad (1)$$

Here, $c_e(T_e)$, $c_{op}(T_{op})$, and $c_{ap}(T_l)$ are the specific heat (units: J cm$^{-2}$ K$^{-1}$) of the electron gas, optical phonon, and acoustic phonon subsystems respectively, with corresponding heat transfer rates between them denoted as $\Gamma_{e\text{-}op}$, $\Gamma_{op\text{-}ap}$, and $\Gamma_{e\text{-}ap}$, and $H$ is the heat injection rate (units: J cm$^{-2}$ s$^{-1}$), which is defined by the absorbed fraction of the incident photon flux. Contribution from the theoretically predicted remote substrate phonons [30,31] is not discussed here, since to our best knowledge their role in hot carrier cooling has not been experimentally observed.

For the case of ultra-short (~50 fs) excitation pulses usually employed for the graphene spectroscopy, the set of rate equations (1) describes the transient temperatures of free carriers and phonons in graphene after the pulse is absorbed, hot carriers are generated, and the ultrafast (~100 fs) carrier-carrier scattering resulted in their equilibrium [26-29]. Therefore, when solving the rate equations, $H$ can be effectively replaced by the initial temperature of the photoexcited electron-hole gas at $t = 0$ [21,22]. The hot carriers relax on the acoustic and optical phonons at $t > 0$ until the system reaches the thermodynamic equilibrium. During the process, high-energy optical phonons relax on acoustic modes of lower energies, increasing the efficiency of carrier-optical phonon cooling pathway.

The specific heat of the electron-hole gas is considered in the most general form [29]:
$c_e(T_e) = d/dT_e \int_{-\infty}^{\infty} \theta(E) f(E) E \, dE$, where $\theta$ is the density of states including spin and valley degrees of freedom, and $f$ is the Fermi-Dirac distribution for electrons or holes. The specific heat function of optical phonons $c_{op}(T_{op})$ is adopted from the work of Lui et al. [11], and that of acoustic phonons is given by $c_{ap}(T_l) \approx 21.6 k_B^3 T_l^2/(2\pi \hbar^2 s^2)$, where $s = 1\times 10^4$ m/s is the sound velocity in graphene [35].

The electron-optical phonon heat transfer rate $\Gamma_{e\text{-}op}$ can be explicitly expressed via the phonon emission rate $R$ as $\Gamma_{e\text{-}op} = R_\Gamma \Omega_\Gamma + R_K \Omega_K$, taking into account both intraband intravalley ($\Omega_\Gamma \approx 0.2$ eV) and intervalley ($\Omega_K \approx 0.16$ eV) electron-phonon scattering [4,36]. The analytical expression for the net emission rate $R_{\Gamma,K}$ includes the phonon distribution function $n_{\Gamma,K}$, and is given by [5,30]:

$$R_{\Gamma,K} = D_{op} \int_{\Omega_{\Gamma,K}}^{\infty} EE' \{f(E)[1-f(E')](1+n_{\Gamma,K}) - f(E')[1-f(E)]n_{\Gamma,K}\} dE.$$

Here, $E' = E - \Omega_{\Gamma,K}$, the electron-phonon coupling efficiency $D_{OP} = 9(dt/db)^2/(\pi \rho \Omega_{\Gamma,K} \hbar^3 v^4)$, where $dt/db \approx 45$ eV/nm [4], $\rho \approx 7.6\times 10^{-7}$ kg/m$^2$ is the density of graphene, and $v = 1.1\times 10^6$ m/s is the speed of free electrons in graphene. The optical phonon subsystem is treated with the Einstein model [11,28] with $n_{\Gamma,K} = [\exp(\Omega_{\Gamma,K}/k_B T_{op}) - 1]^{-1}$, $\Omega_{\Gamma,K}$ is the Einstein frequency. The exact relaxation mechanism of optical phonons on low energy acoustic modes [9,15] is beyond the scope of this paper, and is accounted for via the optical phonon lifetime constant $\tau^{(op)}$, so that $\Gamma_{op\text{-}ap} = c_{op}(T_{op} - T_l)/\tau^{(op)}$ [5] (in this work, we assume the phonon lifetime $\tau^{(op)} = 2$ ps).

The heat transfer rate by the momentum-conserved (normal) electron-acoustic phonon scattering is taken as $\Gamma^{norm}_{e\text{-}ap} = g(T_e - T_l)$, where $g = (D^2 k_B)[15 E_F^4 + 30\pi^2 E_F^2 (k_B T_e)^2 + 7\pi^4 (k_B T_e)^4]/(30\pi \rho \hbar^5 v^6)$, $D$ is the deformation potential [30]. The disorder-assisted supercollision-only electron-acoustic phonon heat transfer rate employed in the SC model [6,21] is given by $\Gamma^{SC}_{e\text{-}ap} = A(T_e^3 - T_l^3)$. Here, $A = \kappa c_e/T_e$; $\kappa = (2/3)[\lambda/(k_F l)](k_B/\hbar)$ is the cooling rate coefficient, being a function of the mean free path $k_F l$ and the electron-acoustic phonon coupling strength $\lambda = [D^2/(\rho s^2)][2 E_F/\pi (\hbar v)^2]$.

Under the setting of Eq. (1), the SC relaxation model corresponds to when $\Gamma_{e\text{-}ap} = \Gamma^{SC}_{e\text{-}ap}$, $\Gamma_{e\text{-}op} = \Gamma_{op\text{-}ap} = 0$; meanwhile the OP model is realized with $\Gamma_{e\text{-}ap} = \Gamma^{norm}_{e\text{-}ap} \ll \Gamma_{e\text{-}op}$, also considering the contribution from weak momentum-conserved electron-acoustic phonon scattering [4,5,30]. Most critically, in the presence of the interface, or the translational defect, we assume that the *interface defect-assisted* electron-acoustic phonon scattering composes an efficient energy dissipation pathway bypassing the momentum conservation via supercollisions [6] (Fig. 1), with an available phonon phase space much larger than the phonon momenta constrained by the Fermi surface for normal collisions. Denoting the interface defect-mediated momentum bypass enhancement factor as $S(T_l)$, we assume the supercollision-driven electron-acoustic phonon heat transfer rate $\Gamma_{e\text{-}ap}$ to be proportional to that of normal collisions: $\Gamma_{e\text{-}ap} = \Gamma^{DP}_{e\text{-}ap} = S(T_l)\Gamma^{norm}_{e\text{-}ap}$. We also expect $S(T_l)$ to be a function of lattice temperature, due to dispersionless acoustic phonons population that follows Bose distribution [10,30]. Meanwhile, we also note that the disorder-attributed mean free path $l$ (70 ~ 140 nm) usually employed in the SC model [6,21,25] is in the same order with the spatial extension of the graphene-metal interface [24,32,33] (40 ~ 150 nm) or graphene $p$-$n$ junction [34] (~300 nm). Consistent experimental reports confirming the negligible contribution from acoustic phonons in uniform graphene [5,7-15,34] also strongly imply the role of interfaces to the unique carrier dynamics derived from the PTE measurements.

Further, we also would like to highlight that, since the factor $S$ directly reflects the effect of the interface defect on the carrier scattering, $S$ can be understood as qualitative measure of the interface features such as for example, impurities concentration, graphene-metal contact quality, and spatial extent of the interface.

## III. RESULTS AND DISCUSSION

### A. Transient PTE photoresponse in graphene

To start, we consider the PTE photocurrent measurements that have been used to support the SC model against the OP model. The instantaneous PTE photocurrent in graphene is given by $i(t) = \beta T_e(t)[T_e(t) - T_l]$, where $\beta$ is proportional to Seebeck coefficient [21-24], and the transient carrier temperature $T_e(t)$ can be obtained by numerically solving the rate equations (1). The time-integrated photocurrent generated by a single pulse with the repetition rate $p$ is calculated as $I_1 = p\int_0^\infty i(t)dt$; and the PTE photocurrent induced by the two successive pulses separated by the delay time $t_d \ll 1/p$ is given by $I_{12}(t_d) = p\int_0^{t_d} i(t)dt + p\int_{t_d}^\infty i(t-t_d)dt$. If $t_d$ is shorter than the photoexcited carrier relaxation time, upon the arrival of the second pulse, a higher value of $c_e(T_e)$ reduces the efficiency of carriers heating relative to that provided by the first pulse. Therefore, two successive pulses generate less PTE photocurrent than a pair of single pulses, and the time-integrated transient photocurrent decay $\Delta I(t_d) = 2I_1 - I_{12}(t_d)$ reflects the transient temperature $T_e(t)$ [21-23].

Taking the parameters of graphene [21] ($E_F$ = 0.1 eV, pulse fluence $F \approx 0.13$ μJ cm$^{-2}$, such that $T_e^{max} \approx 1250$ K), we apply the proposed defect-assisted phonon model (DP model) to numerically obtain the transient $T_e(t)$, and then, the transient $\Delta I(t_d)$. Figure 2(a) shows the experimental data for $\Delta I(t_d)$ at graphene *p-n* junction obtained by Graham *et al*. [21], exhibiting a perfect fit with the suggested DP model (solid) to some extent even better than with the SC model (dashed; with $\kappa = 5.5 \times 10^{-4}$ K$^{-1}$ ps$^{-1}$ [21]). In addition to the transient photocurrent decay $\Delta I(t_d)$, DP model also provides excellent agreement to the measured nonlinear power dependence of the zero-delay photocurrent decay $\Delta I(0)$ measured at $T_l$ = 10 K (Fig. 2(b)) [21], without any further adjustment of free parameters. It is worth noting that, the only fitting parameter in this case is the coefficient $S$ at a given substrate temperature, and the obtained values for $S$ (see Fig. 2(a)) are in reasonable agreement in their order with the cooling power enhancement given by the disorder-based SC model [6]. Independent excellent fits with the recent experimental data by Tielrooij *et al*. [23] (see Supplemental Material [37]), with the same value of $S$ = 180 at room temperature for graphene *p-n* junction and higher $S$ = 220 for the metal contact of narrower spatial extension [24,32,33], also indicate the validity of our approach.

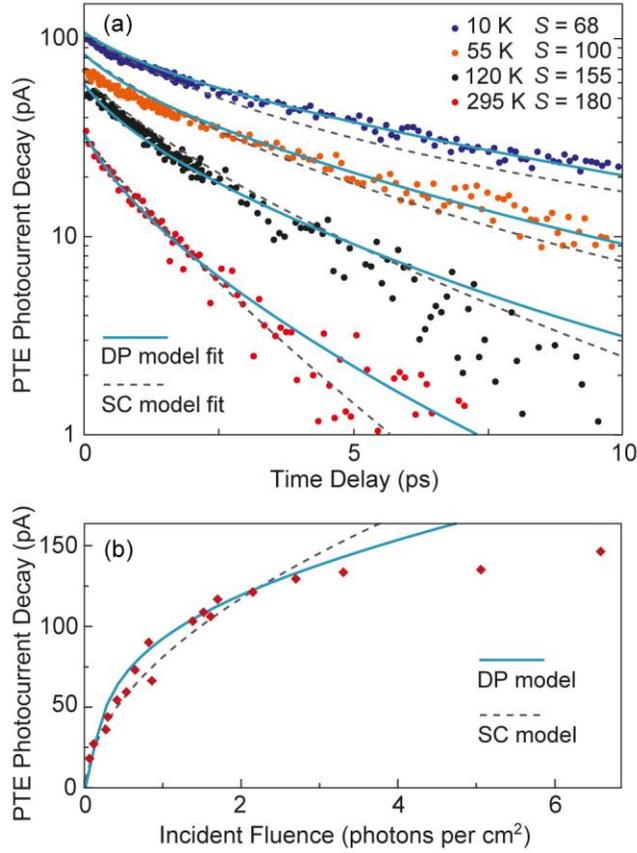

FIG. 2. (a) Transient PTE photocurrent decay measured by Graham *et al*. [21] (dots) at the *p-n* junction in graphene at different lattice temperatures (as denoted); and a fit by the SC (dashed) and DP (solid) models. The DP model considers both the normal electron-optical phonon scattering and the interface defect-assisted supercollisions with acoustic phonons, of its intensity proportional to the momentum-conserved collisions by factor *S*. (b) PTE photocurrent decay at $t_d = 0$ measured by Graham *et al*. [21] (diamonds) at the *p-n* junction in graphene at $T_l = 10$ K as a function of incident fluence; and its calculated value by the DP (solid) and SC (dashed) models.

### B. Cooling power contributions

In order to compare the strength of contributing cooling mechanisms in graphene, we now calculate relaxation rates $\tau^{-1}$ and heat transfer rates $\Gamma$ given by acoustic and optical phonon scattering. From the data by Graham *et al*. [21] (Fig. 2(a)), we obtain the carrier's relaxation rate $\tau_{tot}^{-1} = \tau_{ac}^{-1} + \tau_{op}^{-1}$ at each lattice temperature. As shown in Fig. 3(a), $\tau_{tot}^{-1}$ exhibits the same linear dependence on $T_l$ as the electron-acoustic phonon relaxation rate [30] $\tau_{ac}^{-1} = D^2|E_F|k_BT_l/(4\rho\hbar^3v^2s^2)$. We emphasize that this leads to an electron-optical phonon relaxation rate $\tau_{op}^{-1} \approx (2.4 \text{ ps})^{-1}$ independent of lattice temperature. This is in good agreement with previous experimental report made for graphene with non-zero Fermi level [14], and also in line with the theoretical prediction for $T_l$-independent $\Gamma_{e\text{-op}}$ (Fig. 3(b), dashed lines). Based on the agreement of obtained $\tau_{op}^{-1} \approx (2.4 \text{ ps})^{-1}$ to the previous experiments [14], and noting that

$\Gamma_{\text{e-op}}$ and $\Gamma^{DP}_{\text{e-ap}}$ are of the same order over the most of the hot carriers' temperatures (Fig. 3(b)), we thereby conclude the presence of the two concurrent and equally efficient relaxation mechanisms: the normal collisions with optical phonons and the interface-assisted supercollisions with acoustic phonons. The cases of $\Gamma^{DP}_{\text{e-ap}}$ dominated or $\Gamma_{\text{e-op}}$ dominated cooling are also addressed in Supplemental Material [37], and are in agreement with the proposed model.

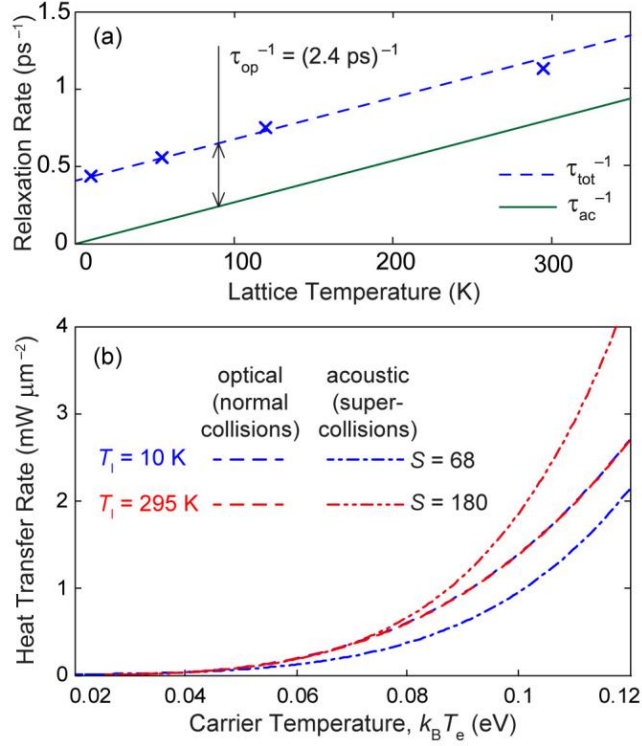

FIG. 3. (a) Hot carriers' relaxation rate $\tau_{\text{tot}}^{-1}$ (cross marks) obtained from the experimental data by Graham *et al*. [21], its linear fit (dashed line), and theoretical relaxation rate $\tau_{\text{ac}}^{-1}$ (solid line) calculated for the normal collisions with acoustic phonons. The distance between the two corresponds to the temperature-independent electron-optical phonons relaxation rate $\tau_{\text{op}} \approx 2.4$ ps, in agreement with the experimental report [14]. (b) Analytical solution for the net heat transfer rate from the hot carriers to the optical phonon bath ($\Gamma_{\text{e-op}}$; dashed), and that to the acoustic phonon bath via the interface defect-assisted supercollisions ($\Gamma^{DP}_{\text{e-ap}}$; solid); under the assumption of $T_{\text{op}} = T_{\text{ac}} = T_l = 295$ K (red), and 10 K (blue); $E_F = 0.1$ eV.

### C. Nonlinear response at low excitation power

Finally, we test the proposed DP model against the recently observed nonlinear PTE photoresponse under weak excitation [38], a phenomenon that is not predicted by either OP or SC models. Applying the DP model, generated by a single pulse integrated photocurrent $I$ is calculated at low excitation pulse power (up to 6 µW, which corresponds to a maximum carrier temperature of 760 K). Figure 4 shows the calculated PTE photocurrent behavior in good agreement with the experimental observations: stronger and sublinear photocurrent response at low doping ($E_F$ = 0.2 eV, Fig. 4(a)), versus weaker and superlinear response at high doping ($E_F$ = 0.5 eV, Fig. 4(b)). This agreement with the experimental data [38] by using the same parameters as in Fig. 2 [21], except for the changes in graphene doping level, confirms the theoretical consistency and versatility in the application of our model.

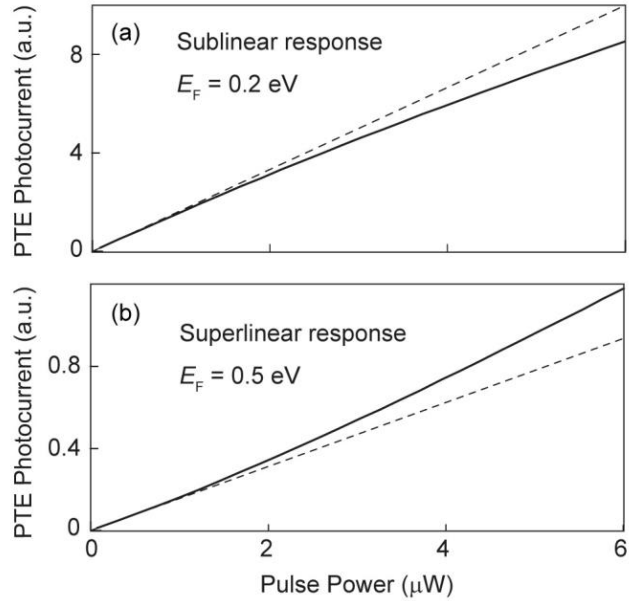

FIG. 4. DP-model calculated PTE photocurrent (solid line) at graphene *p-n* junction by a single laser pulse of low intensity: assuming (a) moderate ($E_F$ = 0.2 eV) and (b) high doping ($E_F$ = 0.5 eV) in graphene. Dashed line illustrates an asymptote to the photocurrent curve at zero. System parameters are the same as used to obtain the fit for data of Ref. [21] (in Fig. 2(a), at room temperature).

## IV. CONCLUSION

To summarize, we propose a new model for photoexcited carrier cooling in graphene, considering both electron-optical phonon and electron-acoustic phonon (supercollision) scattering. Motivated by the experimentally observed inconsistency between the interface-driven and uniform graphene photoresponse, we make a reasonable assumption addressing the role of the localized interface defect, as a momentum bypass route initializing the electron-acoustic phonon supercollisions. The application of proposed model to the transient and nonlinear PTE photoresponse of graphene *p-n* junction and graphene-metal contact demonstrates perfect fits to the experimental data; suggesting that the photocurrent measurement reflects the effect of the interface rather than the intrinsic properties of graphene. Furthermore, based on the calculated heat transfer rates, we conclude that the contribution of the acoustic phonons to the cooling, otherwise negligible for uniform graphene layer, is enhanced in the presence of the localized interface defect (supercollisions), and becomes as efficient as the inherent optical phonon-mediated cooling in graphene. Highlighting previously overlooked effect of the interface for the cooling dynamics, and providing a unified theoretical ground for the ultrafast photoresponse in graphene, we expect our results to propel the practical investigation of graphene photoresponse in general, and to assist the study of hot carrier dynamics in particular.

## ACKNOWLWDGEMENTS

Authors wish to thank Dr. Sunkyu Yu for valuable discussions, and acknowledge support from the National Research Foundation of Korea (NRF) through the Global Frontier Program (GFP, 2014M3A6B3063708) and the Global Research Laboratory Program (GRL, K20815000003), all funded by the Ministry of Science, ICT & Future Planning of the Korean government.

# Supplemental Material

# Interface defect-assisted phonon scattering of hot carriers in graphene

Sergey G. Menabde,[1] Hyunwoo Cho,[1] and Namkyoo Park[1,*]

[1]*Dept. of Electrical and Computer Engineering, Seoul National University, Seoul, 08826, Korea*

*nkpark@snu.ac.kr

## 1. Fit to recent experiments at *p-n* junction and metal contact

Figure S1 demonstrates the fit provided by our defect-assisted phonon model to the experimental results obtained by Tielrooij *et al*. [22,23] for the PTE photocurrent decay at graphene *p-n* junction and graphene-metal contact at room temperature. The same value of $S = 180$ is used to fit the data from the *p-n* junction as used for the results in Ref. [21], reflecting the same nature of the interface defect. Moderately higher $S = 220$ is used to fit the photocurrent decay at the graphene-metal contact, indicating sharper interface profile in this case (see the main text). Absorbed fluence from the excitation pulse is set to $F = 0.02$ μJ cm$^{-2}$, which corresponds to the reported values of the excitation power [21]. Fermi level in both cases is assumed $E_F = 0.13$ eV which is typical for the unprotected graphene on a substrate [6,21]. Although the fitting value for the Fermi level used in Ref. [22] is $E_F = 0.2$ eV, we believe that the conventional value around 0.1 eV is closer to the considered case of globally undoped single layer graphene under the given experimental conditions.

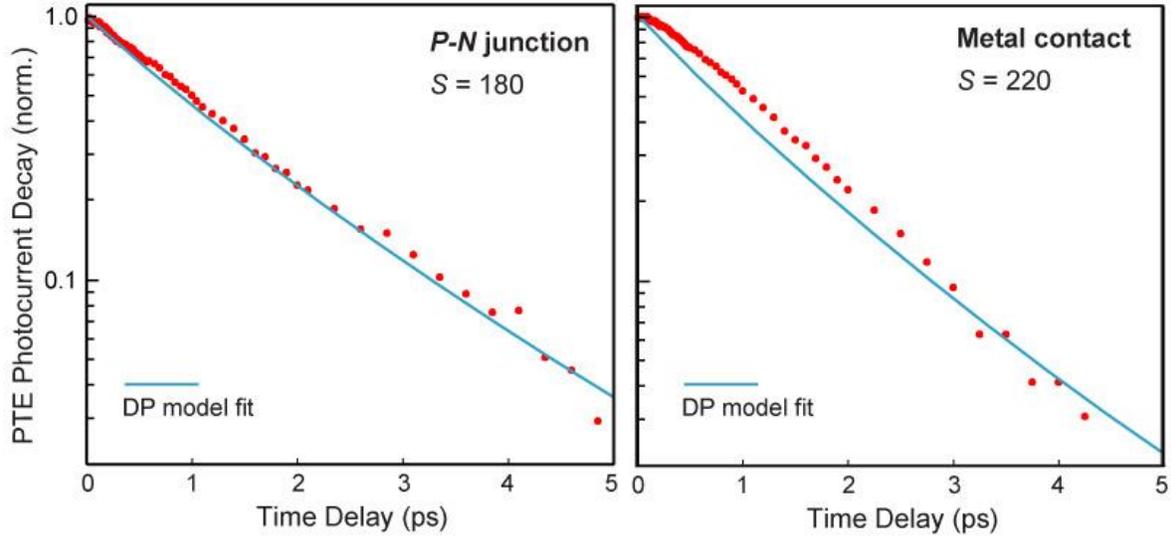

FIG. S1. Experimental values for the PTE photocurrent decay (dots) published by Tielrooij *et al.* in Ref. [23] at graphene *p-n* junction (left) and metal contact (right), at room temperature. The defect-assisted phonon model provides a perfect fit (solid line) for the *p-n* junction, as well as a good fit for the metal contact. The same value of $S = 180$ is used to fit the similar experimental results by Graham *et al.* [21] for the PTE current decay generated at the *p-n* junction at room temperature. Fitting the metal contact requires a bit higher *S*. Other model parameters are the same in both cases: absorbed fluence $F = 0.02$ μJ cm$^{-2}$ corresponds to the reported values of the excitation power, $E_F = 0.13$ eV, and $T_l = 295$ K.

## 2. Electron-phonon heat transfer rates via optical and acoustic modes

Figure S2 demonstrates theoretical electron-phonon heat transfer rates Γ, considering phonon bath temperature to be constant and equal to the lattice temperature $T_l$; Fermi level $E_F = 0.1$ eV.

As shown in Fig. S2, the heat transfer rate via the electron-optical phonon scattering ($\Gamma_{e\text{-}op}$; dashed lines) dominates the cooling process, exceeding the heat transfer rate by normal collisions with acoustic phonons ($\Gamma^{norm}_{e\text{-}ap}$; thin solid lines) by the two orders of magnitude. This case corresponds to the conventional OP model [7-15], or when $S \approx 0$. Furthermore, $\Gamma_{e\text{-}op}$ practically does not depend on lattice temperature for hot carriers, as also noted in the main text. In the presence of the interface defect however, i.e. when $S > 0$, the contribution from the interface defect-assisted electron-acoustic phonon supercollisions $\Gamma^{DP}_{e\text{-}ap} = S\Gamma^{norm}_{e\text{-}ap}$ (thick solid lines) is of the same order as $\Gamma_{e\text{-}op}$. The comparable values of $\Gamma_{e\text{-}op}$ and $\Gamma^{DP}_{e\text{-}ap}$ are observed for the whole temperature range when the substrate temperature

$T_l = 295$ K. Meanwhile, for lower lattice (phonon) temperature, we predict a domination of interface defect-assisted scattering due to the energy bottleneck manifestation for the high energy optical phonons, when their emission rate $R_{\Gamma,K}$ is significantly suppressed by sharp distribution functions in the system. This case directly refers to the experimental observations by Graham *et al.* [21,25], when the efficient cooling of carriers at low temperature has been associated with the disorder-driven supercollisions (i.e. the SC model [6]), instead of the enhanced scattering due to the localized defect of the photocurrent generating *p-n* junction.

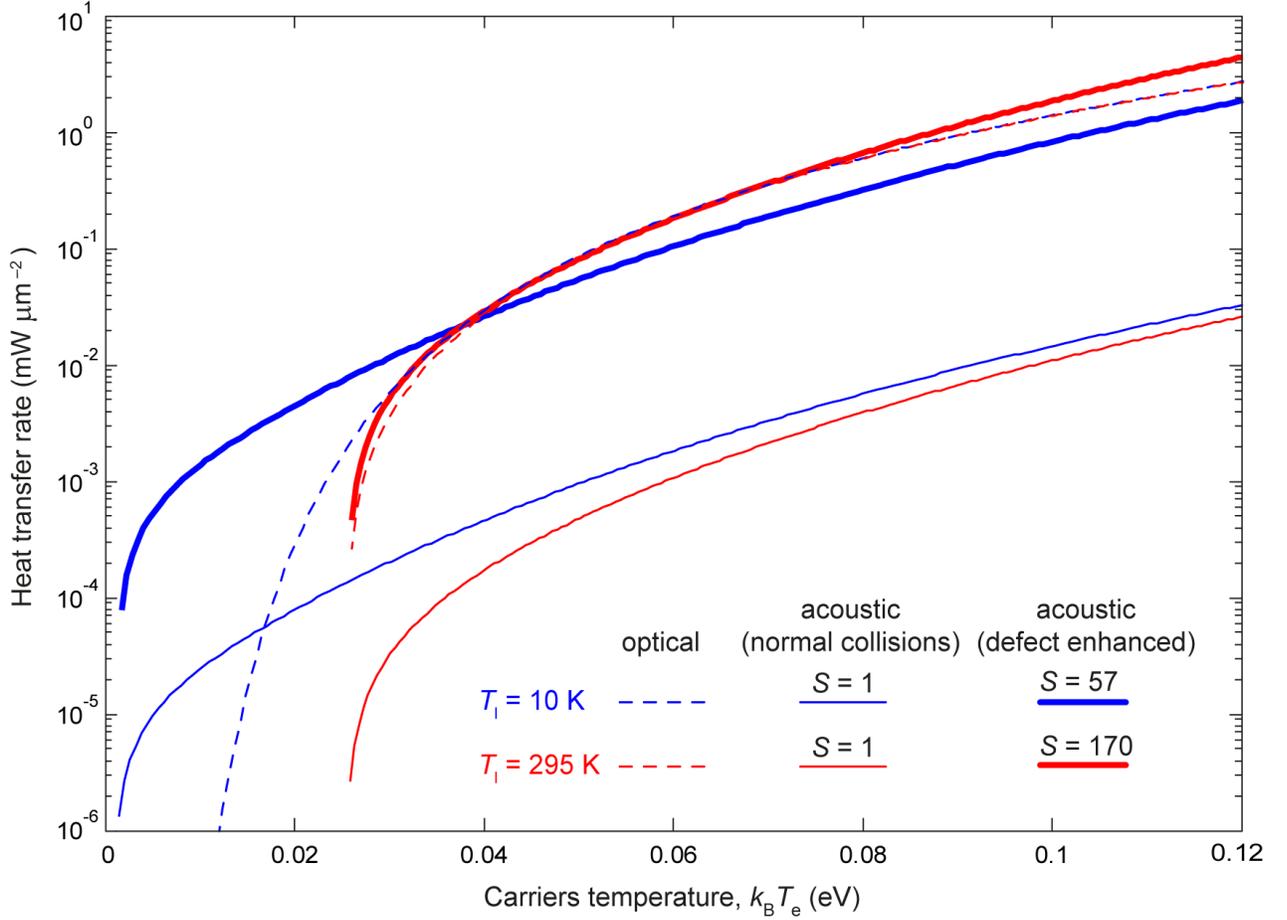

FIG. S2. Theoretical net heat transfer rates $\Gamma$ for the electron-phonon cooling as a function of electronic temperature; obtained for the normal scattering by the optical ($\Gamma_{e\text{-op}}$; dashed) and acoustic ($\Gamma^{norm}_{e\text{-ap}}$; thin solid) phonons, as well as the defect-enhanced scattering by acoustic phonons ($\Gamma^{DP}_{e\text{-ap}}$; thick solid). Results are obtained for constant phonon and lattice temperatures 10 K (blue) and 295 K (red).